\documentclass[aps,prl,twocolumn,amsmath,amssymb,nofootinbib,showpacs,superscriptaddress]{revtex4}
\usepackage[english]{babel}
\usepackage{latexsym}
\usepackage{graphics}
\usepackage{epsfig}
\usepackage{color}

\def\bea{\begin{eqnarray}}
\def\eea{\end{eqnarray}}
\def\be{\begin{equation}}
\def\ee{\end{equation}}

\begin{document}
\title{Parametric Excitation of a 1D Gas in Integrable and Nonintegrable Cases}

\author{M. Colom\'e-Tatch\'e}
\affiliation{\mbox{Institut f\"ur Theoretische Physik, Leibniz Universit\"at Hannover, Appelstr. 2 D-30167, Hannover, Germany}}

\author{D.S.~Petrov}
\affiliation{\mbox{Laboratoire de Physique Th\'{e}orique et Mod\`{e}les Statistiques, CNRS,
Universit\'{e} Paris Sud, 91405 Orsay, France}}
\affiliation{Russian Research Center Kurchatov Institute,
Kurchatov Square, 123182 Moscow, Russia}

\date{\today}

\begin{abstract}
We study the response of a highly excited 1D gas with pointlike interactions to a periodic modulation of the coupling constant. We calculate the corresponding dynamic structure factors and show that their low-frequency behavior differs dramatically for integrable and nonintegrable models. Nonintegrable systems are sensitive to excitations with frequencies as low as the mean level spacing, whereas much higher frequencies are required to excite an integrable system. This effect can be used as a probe of integrability for mesoscopic 1D systems and can be observed experimentally by measuring the heating rate of a parametrically excited gas.
\end{abstract}
\maketitle

The field of ultracold gases has progressed enormously toward obtaining quantum systems with desired densities and types of constituent atoms, trapping geometries, and well controlled interparticle interactions \cite{Bloch-Dalibard-Zwerger}. In particular, by using either an optical potential or large magnetic field gradients, one can confine the motion of atoms to one dimension and create interacting 1D gases of bosons \cite{1DBosons} and fermions \cite{CI-molecule,Randy1DFermions}, which can be described by the integrable Lieb-Liniger \cite{Lieb-Liniger} and Yang-Gaudin \cite{CNYang,Gaudin} models respectively. The purity and isolation of such systems from the environment makes them ideal candidates for studies of fundamental differences between integrable and nonintegrable many-body dynamics. A pioneering experiment on this subject has been performed recently by Kinoshita and co-workers \cite{Kinoshita}. They have shown that a 1D Bose gas initially prepared in a highly excited state does not equilibrate in the lifetime of the experiment, whereas essentially the same system with a weaker 1D confinement thermalizes much faster.

How do we decide whether a system is integrable or not? Let us put aside strict mathematical definitions of quantum integrability \cite{Integrability} and look at the problem phenomenologically. What measurement should we perform on a system in order to conclude on its integrability? The field of quantum chaos suggests to look at its spectral statistics \cite{Percival}. If energy levels are not correlated [the nearest neighbor spacing distribution (NNSD) is Poissonian], we are dealing with an integrable or regular system \cite{Berry1}. If, in contrast, levels repel each other, the system is not integrable \cite{Porter,BohigasGiannoniSchmit}. Spectral properties were extensively studied for various systems \cite{Percival,Berry1,Porter,BohigasGiannoniSchmit}, and, in particular, for strongly correlated condensed matter models (see \cite{ManyBodySpecStat} for early work).

Another signature of integrability is the localization of eigenstates of a regular system in a certain physically meaningful basis \cite{Berry2}, which suggests the dynamical probe of integrability: if one creates an excited initial state localized in this basis, it will stay localized during the temporal evolution. In fact, the absence of thermalization in the experiment \cite{Kinoshita} can be regarded as a consequence of the localization of the Lieb-Liniger eigenstates in momentum space.

In this Letter we compare responses of highly excited integrable and nonintegrable systems to an external time-dependent perturbation. We explore the idea that a perturbation localized in the same space as the eigenstates of the integrable system probes its local density of states, whereas in the nonintegrable case the states are delocalized, and the perturbation, no matter localized or not, couples all of them. Considering two 1D models on a ring we demonstrate that integrable systems can be much more stable with respect to slow variation of their Hamiltonian than nonintegrable ones. Namely, we consider the model of a single mobile impurity in a Fermi gas and the Lieb-Liniger model, and study their response to a periodic modulation of the coupling constant. This perturbation is localized in the many-body momentum space as it only changes the relative momentum of an atom pair. We show that the nonintegrable system is sensitive to excitations with frequencies as low as the many-body mean level spacing, which is exponentially small, whereas the threshold frequency in the integrable case is much larger and scales polynomially with the system size.

Consider $N$ atoms with short range interactions in a quasi-1D ring-shaped trap of circumference $L=1$. If the atomic kinetic energies are smaller than the level spacing in the direction of tight confinement, the system can be envisioned as a 1D gas on a ring with the Hamiltonian
\be
\label{eq:GeneralHamiltonian}
H=-\sum_{i=1}^{N}\frac{1}{2m_i}\frac{\partial^2}{\partial x_i^2}+\sum_{i<j}g^{ij}\delta(x_i-x_j),
\ee
where $m_i$ and $x_i$ are the masses and coordinates of the atoms. The 1D coupling constants $g^{ij}$ depend on the parameters of the 3D interatomic interactions as well as on the strength of the tight confinement \cite{Olshanii}. Accordingly, by changing these parameters in time one can study the response of the effective 1D system to variations of $g^{ij}$.

Assume that the system is initially prepared in an eigenstate of the Hamiltonian (\ref{eq:GeneralHamiltonian}) with eigenvalue $\varepsilon_\nu$ and eigenfunction $\psi_\nu$, and consider the weak periodic modulation  $g^{ij}(t) = g^{ij}+2\delta g^{ij}\cos(\omega t)$. Then, in the linear response regime the probability to remain in the state $\nu$ decreases with the rate $\Omega=2 \pi \left[ S(\varepsilon_{\nu},\omega)+S(\varepsilon_{\nu},-\omega)\right]$, where the dynamic structure factor equals
\be
\label{eq:DSF}
S(\varepsilon_\nu,\omega)=\sum_{\eta}\delta(\omega - \varepsilon_\eta + \varepsilon_\nu)| \langle\psi_{\eta}|F|\psi_{\nu}\rangle |^2,
\ee
and where $F=\sum_{i<j}\delta g^{ij}\delta(x_i-x_j)$. Exciting an ensemble of systems (\ref{eq:GeneralHamiltonian}) leads to a diffusion of the population in energy space with diffusion constant $2\pi \omega^2 S(\varepsilon,\omega)$ resulting in detectable changes of the total energy and entropy.

The asymptotic behavior of $S(\varepsilon,\omega)$ at small $\omega$ gives the dissipative part of the response of the system to a slow variation of its Hamiltonian and, therefore, measures the degree at which this variation can be assumed adiabatic. For complex systems it is believed that statistical properties of eigenstates, eigenvalues, and matrix elements of a perturbation are well described by the random matrix theory \cite{Wilkinson}. If both $H$ and $F$ were independent random matrices drawn from the Gaussian Orthogonal Ensemble, the average of $S(\varepsilon_\nu,\omega)$ over an energy interval larger than the mean level spacing $D(\varepsilon)$ would be independent of $\varepsilon$ and $\omega$ \cite{rem}. Here we show that this low-frequency behavior strongly depends on whether we consider integrable or nonintegrable systems.

Let us consider the model of a single mobile impurity interacting with a gas of identical ideal fermions. In this case the parameters of (\ref{eq:GeneralHamiltonian}) are $m_1=...=m_{N-1}=1$, $m_N=M$, and $g^{iN}=g$. It is convenient to work in momentum space introducing the Fourier transform
\begin{equation}\nonumber
\psi(x_1,...,x_{N})=\sum_{p_1,...,p_{N}}\psi(p_1,...,p_{N})e^{-ip_1 x_1...-ip_N x_N},
\end{equation}
where all the momenta are integer multiples of $2\pi$. The total momentum $Q$ is conserved and keeping in mind that $p_N=Q-\sum_{i=1}^{N-1}p_i$ we omit the argument $p_N$ in the wavefunction $\psi$, which now becomes antisymmetric in all of its arguments. The Schr\"odinger equation then reads
\begin{equation}\label{MomentumSchroedinger}
\left(\sum_{i=1}^{N}\frac{p_i^2}{2m_i}-E\right)\psi=-g\sum_{i=1}^{N-1}\sum_{p'_i}\psi (p_1,...,p'_i,...,p_{N-1}).
\end{equation}
Let us introduce an auxiliary function
\begin{equation}\label{alpha}
\alpha(p_1,...,p_{N-2})=-\sum \text{\raisebox{-5pt}{$\scriptstyle p_{N-1}$}}\psi(p_1,...,p_{N-1})
\end{equation}
and using the antisymmetry of $\psi$ rewrite the rhs of Eq.~(\ref{MomentumSchroedinger}) in the form $g[\alpha-\sum_{i=1}^{N-2}\alpha(...,p_{i-1},p_{N-1},p_{i+1},...)]$ (Hereafter, for normally ordered arguments we use the shortcut $\alpha\equiv\alpha(p_1,...,p_{N-2})$). We then solve Eq.~(\ref{MomentumSchroedinger}) with respect to $\psi$ and substitute the result into the definition of $\alpha$ (\ref{alpha}) obtaining the equation
\begin{equation}\label{TheEquation}
\left[\frac{1}{2\mu g}+\frac{(1/2\kappa)\sin \kappa}{\cos \kappa -\cos \xi}\right]\alpha=\sum_q\sum_{i=1}^{N-2}\frac{\alpha(...,p_{i-1},q,p_{i+1},...)}{(q-\xi)^2-\kappa^2},
\end{equation}
where $\mu=M/(M+1)$, $\xi=(\mu/M)(Q-\sum_{i=1}^{N-2}p_i)$, and $\kappa^2=2\mu E-\mu\sum_{i=1}^{N-2}p_i^2-M\xi^2$. 

By solving Eq.~(\ref{TheEquation}) with respect to $E$ and $\alpha$ we determine the eigenenergies $\varepsilon_\nu$ and the ``reduced'' eigenfunctions $\alpha_\nu$. We then calculate the dynamic structure factor (\ref{eq:DSF}) for $F=\delta g\sum_{i=1}^{N-1} \delta (x_i-x_N)$ by using the relation $\langle \psi_{\eta}|\delta (x_i-x_N)|\psi_{\nu}\rangle= \sum_{p_1,...,p_{N-2}}\alpha_\eta^*\alpha_\nu$. Note that Eqs.~(\ref{MomentumSchroedinger}) and (\ref{TheEquation}) conserve parity (simultaneous sign change of all $p_i$). The corresponding even and odd excitation branches are not coupled by $F$. They have the same density of states and contribute equally to Eq.~(\ref{eq:DSF}).

\begin{figure*}[hptb]
\begin{center}
\includegraphics[width=1.6\columnwidth]{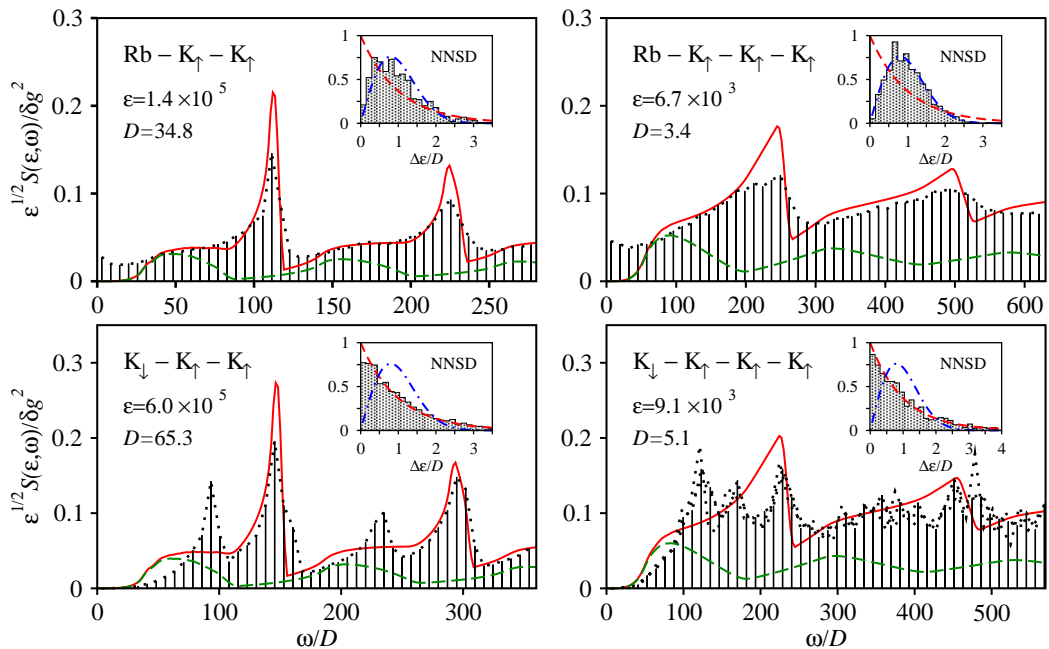}
\caption{(color online) Dotted lines present dynamic structure factors for nonintegrable (upper panel) and integrable cases (lower panel). Red solid and green dashed lines are the result of the binary approximation (see text). Insets show the nearest neighbor spacing distributions for the ensembles of energy levels used to calculate $S$ in the corresponding parent graphs. Red dashed and blue dash-dotted lines stand, respectively, for the Poissonian and Wigner-Dyson distributions.}
\vspace{-0.5cm}
\label{fig:DSF}
\end{center}
\end{figure*}

We have performed an extensive numerical analysis of this model for $N=3,4$ in a wide range of energies, coupling constants, and for different $M$. We accurately calculate up to $10^4$ excited levels. In the integrable case, $M=1$, we determine $\varepsilon_\nu$ and $\psi_\nu$ from the known Bethe-ansatz solution \cite{McGuire} and check that both approaches give the same result. In Fig.~\ref{fig:DSF} we plot $S(\varepsilon,\omega)$ for four cases: The upper panels stand for the nonintegrable case with $N=3$ (left) and $N=4$ (right) with $M=m_{87{\rm Rb}}/m_{40{\rm K}}$. The lower panels show the integrable case.

Dotted lines filled vertically to the $x$-axis present the average of $S(\varepsilon_\nu,\omega)\sqrt{\varepsilon}/\delta g^2$ over several hundreds of states with odd parity and zero total momentum in the interval $0.95\varepsilon<\varepsilon_\nu<1.05\varepsilon$. All four presented cases correspond to the same value of the interaction parameter $\gamma=\langle k_{rel}^2\rangle a_1^2\approx 0.5$, where $\langle k_{rel}^2\rangle=2\mu\varepsilon/(N-1)$ is the average square of the relative two-body momentum, and $a_1=1/\mu g$ is the 1D scattering length. We check that simultaneous variation of $g$ and $\varepsilon$ does not change the quantity $S(\varepsilon,\omega/\sqrt{\varepsilon})\sqrt{\varepsilon}$ as long as $\gamma$ stays constant. The lower right panel shows three dotted lines obtained for $\varepsilon=$6.9, 9.1, and 14.2 [$\times 10^3$] with the same $\gamma$. With rescaled $x$-axes they collapse to a single curve. The noise is due to the averaging over finite number of states and is uncorrelated between different curves. We measure $\omega$ in units of the mean level spacing $D(\varepsilon)$, so the labelling of the horizontal axis holds only for the curve $\varepsilon=9.1\times 10^3$. 

In all our calculations (not only in Fig.~\ref{fig:DSF}) we observe that the low-frequency behavior is universal: In nonintegrable cases we {\it always} see that $S(\omega)$ tends to a finite constant when $\omega\sim D$, which is consistent with the random matrix theory and with the fact that all states are coupled by the perturbation. In contrast, all integrable cases are {\it always} characterized by a strong suppression of $S(\omega)$ already for quite large $\omega/D$, which means that the perturbation does not couple states with close energies. This is the main result of our numerical experiment.

In order to understand this behavior and the peaks at higher frequencies we use the cluster expansion up to binary terms, i.e. we assume that only pairs of atoms can be excited at a time, the remaining particles being non-interacting spectators. The red solid lines in Fig.~\ref{fig:DSF} correspond to the quantity $\bar S(\varepsilon,\omega) = (N-1)D(\varepsilon) \int_0^\varepsilon \bar \rho(\varepsilon-E) S_2(E,\omega) \rho_2(E) {\rm d} E$, where the prefactor $N-1$ is the number of interacting pairs, $S_{2}(E,\omega)$ and $\rho_2(E)$ are, respectively, the dynamic structure factor and the density of states for an atom pair, and $\bar \rho(\varepsilon-E)$ is the ideal-gas density of states for the remaining atoms.

For highly excited two-body states with energies $\approx E$ and center-of-mass momenta $Q$ Eq.~(\ref{TheEquation}) gives two excitation branches: $\kappa_{n,\pm}=2\pi(n-1/4)+\phi\pm\Delta$, where $\phi=-\arctan (a_1\kappa)$, $\kappa^2=2\mu E-M\xi^2$, $\xi=-\mu Q/M$, and $\Delta=\arccos(\sin\phi\cos\xi)$. The sum in Eq.~(\ref{eq:DSF}) splits into intra- and interbranch excitations:
\begin{equation}\label{S2}
S_2(E,\omega)=\sum_{Q,q} A_\pm \delta (\omega-q\kappa/\mu)+B \delta [\omega-(\kappa/\mu)(q\mp 2\Delta)],
\end{equation}
where $Q$ and $q$ are integer multiples of $2\pi$, $A_\pm=\delta g^2\sin^4\phi (1\mp\cos\xi/\sqrt{1+a_1^2\kappa^2\sin^2\xi})^2$, $B=\delta g^2\sin^2\phi\tan^2\phi\sin^2\xi/(1+a_1^2\kappa^2\sin^2\xi)$, and the sign corresponds to the choice of the initial state. 

For unequal masses $\xi$ is incommensurate with $\pi$, and one can change summation over $Q$ by integration. The first term in Eq.~(\ref{S2}) leads to strong peaks of $\bar S$ at frequencies which are integer multiples of $2\pi\sqrt{2\varepsilon/\mu}$, whereas the second term gives lower and wider interbranch lobes shown separately as green dashed lines. The widening is due to the averaging of $\Delta$ when integrating over $Q$.

In the case of equal masses the quantization of the center-of-mass motion imposes $\xi=Q/2=\pi m$. Then, the two branches correspond to symmetric and antisymmetric two-body states, the interbranch excitations are not possible ($B\equiv 0$) and only the symmetric branch is sensitive to the variation of the coupling constant [$A_\pm=\sin^4\phi(1\mp(-1)^m)$]. This completely ignores the interbranch peaks and strongly overestimates the intrabranch ones, contrary to the numerics. Better agreement is obtained by assuming the continuum uniform distribution of $Q$ as in the nonintegrable cases -- all red solid curves in Fig.~\ref{fig:DSF} are obtained by integrating over $Q$. The proper distribution of $Q$ and the nature of the interbranch peaks is beyond the two-body physics. Yet, the binary approximation qualitatively explains the low-frequency suppression of $S$ in the integrable case.

Beyond the case $\gamma\approx 0.5$ presented in Fig.~\ref{fig:DSF} we calculate $S(\varepsilon,\omega)$ for $\gamma$ varying from $\sim 10^{-2}$ to $\sim 10^{2}$. Comparing with Fig.~\ref{fig:DSF}, for stronger interactions (smaller $\gamma$) the interbranch lobes turn into sharper peaks consistent with the binary approximation (in this case $\Delta$ is always close to $\pi/2$). For larger $\gamma$ the peaks smoothen and $S(\varepsilon,\omega)$ is well approximated by $\bar S(\varepsilon,\omega)$ except for small $\omega$, where $S$ tends to a constant in the nonintegrable case. We also calculate $S(\varepsilon,\omega)$ for $M=m_{40{\rm K}}/m_{6{\rm Li}}$ and $M=m_{6{\rm Li}}/m_{40{\rm K}}$ and see no qualitative deviations from Fig.~\ref{fig:DSF}.

\begin{figure}[hptb]
\begin{center}
\includegraphics[width=0.8\columnwidth,clip]{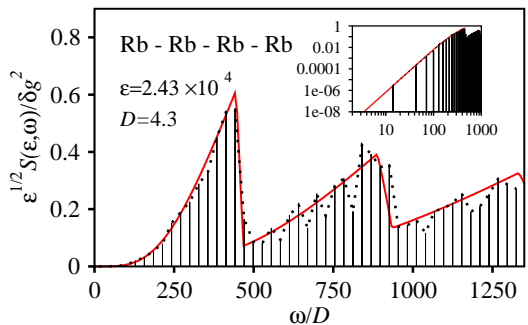}
\caption{(color online) Dynamic structure factor for the Lieb-Liniger model (dotted line) and the binary approximation (red solid line). Inset shows the same data on log-log scale.}
\vspace{-0.6cm}
\label{fig:LL}
\end{center}
\end{figure}

We have also performed an extensive analysis of the bosonic Lieb-Liniger model given by Eq.~(\ref{eq:GeneralHamiltonian}) with $m_i\equiv 1$ and $g^{ij}\equiv g$. We find that here the two-body approximation works perfectly well (with $Q=2\pi m$) for all coupling constants and particle numbers ($N\leq 4$) that we have considered. Figure~\ref{fig:LL} is a representative plot of $S(\varepsilon,\omega)$ for a four-boson system. We observe that at low frequencies $S\propto\omega^4$ (see inset in Fig.~\ref{fig:LL}). This is, actually, a manifestation of the 1D fermionization as small frequencies correspond to small relative momenta $\kappa=\mu\omega/2\pi$ [see Eq.~(\ref{S2})]. The probability to find two atoms at small distances drops as $\kappa^2 a_1^2$, which directly transfers to the matrix elements of the perturbation and eventually leads to the $\kappa^4$-behavior of the coefficient $A_\pm$ in Eq.~(\ref{S2}).

Apart from this $\omega^4$-suppression at small frequencies the binary approximation prohibits excitations with $\omega\lesssim 1/L^2$, which is the minimal level spacing for the two-body problem. Assuming that this result holds for large $N\propto L$, we can conclude that the original state of the system is preserved if the perturbation is slow polynomially in the system size. This is consistent with the statement on the absence of adiabaticity in 1D systems in the thermodynamic limit \cite{Polkovnikov}. Note, however, the distinction between polynomially long timescales for integrable systems and exponentially long ones in nonintegrable cases, which is interesting from the quantum computing perspective.

Our particle number is limited mostly by the complexity of the nonintegrable model. There is still some room for increase, but we expect no qualitative change of the system behavior, at least at small $\omega$ where the results are consistent with the random matrix theory. In contrast, going to larger $N$ in the integrable cases and developing a smarter approach for calculating matrix elements seems to be an interesting theoretical project. Caux and Calabrese have recently proposed an efficient numerical algorithm for calculating correlation functions based on the algebraic Bethe ansatz \cite{Calabrese}.

We thank E. Bogomolny, O. Bohigas, R. Dubertrand, M. Zvonarev, M. Olshanii, B. Altshuler, and P. Calabrese for fruitful discussions and acknowledge support by the IFRAF Institute, by ANR (grant 08-BLAN-65), by the EuroQUAM-FerMix program, by the Centre for Quantum Engineering and Space-Time Research QUEST, and by the Russian Foundation for Fundamental Research.


\begin{thebibliography}{99}
\bibitem{rem} More precisely, for $\omega\ll D$ this average is proportional to $\omega$, which follows from the statistics of Landau-Zener crossings and can be derived by the method of \cite{Wilkinson}.
\end{thebibliography}
\end{document}